\documentclass[prd,twocolumn,nofootinbib]{revtex4}
\usepackage{graphics}
\usepackage{SIunits}
\usepackage{amsmath}
\usepackage{tensind}

\tensordelimiter{:}
\tensorformat{}
\whenindex{,}{\,,\,}{\finishdots}
\whenindex{;}{\,;\,}{\finishdots}
\whenindex{|}{\,|\,}{\finishdots}

\addunit{\jansky}{Jy}
\addunit{\Bel}{B}

\DeclareMathOperator{\diag}{diag}
\newcommand*{\diff}{\,\mathrm{d}}
\newcommand*{\ee}{\mathrm{e}}
\newcommand*{\aye}{\mathrm{i}}
\newcommand{\radio}[1]{#1_{\text{r}}}
\newcommand{\gw}[1]{#1_{\text{gw}}}
\newcommand*{\omegar}{\radio{\omega}}
\newcommand*{\omegagw}{\gw{\omega}}
\newcommand*{\magnitude}[1]{\left| #1 \right|}
\newcommand*{\boltzman}{k_{\text{B}}}
\newcommand*{\Snu}{S_{\nu}}
\newcommand*{\Tsys}{T_{\text{sys}}}
\newcommand*{\Tant}{T_{\text{A}}}
\newcommand*{\Aeff}{A_{\text{eff}}}
\newcommand*{\snr}{\text{SNR}}
\newcommand{\rms}[1]{{#1}_{\text{rms}}}
\newcommand*{\Phieff}{\Phi_{\text{eff}}}
\newcommand{\sol}[1]{{#1}_{\odot}}
\newcommand*{\efficiency}{\epsilon}

\begin{document}

\title{The Response of a Two-Element Radio Interferometer to Gravitational
Waves}

\author{Kipp Cannon}
\affiliation{Department of Physics, University of Alberta, Edmonton,
Alberta, Canada, T6G 2J1}
\email{kcannon@phys.ualberta.ca}
\date{November 19, 2003}

\begin{abstract}
This document presents a ray-optics analysis of the response of a
two-element radio interferometer to the presence of a plane gravitational
wave.  A general expression for the differential phase observed between the
two receiving stations as a result of an arbitrary gravitational wave is
determined, as well as the specific responses to monochromatic and black
hole ring-down waveforms.  Finally, the possibility of gravitational wave
detection via this mechanism is discussed in the context of interferometer
noise.
\end{abstract}


\maketitle

\section{Introduction}

A gravitational wave is a perturbation to the geometry of space that
propagates through space at the speed of light.  It's effect is to stretch
and shrink the distances between objects in the directions transverse to
its direction of travel.  The distortion is volume-preserving, so while it
is shrinking distances in one direction, it stretches them in another.

The stretching and shrinking effect is not a co-ordinate artifact.  A
measuring stick used to monitor the distance between two objects
freely-falling in the presence of a gravitational wave will measure one
separation distance at one time and a different separation later.  The
stretching and shrinking is also linear:  given the change in distance
between two objects separated by a distance \(L\), the change in distance
between two objects separated by \(2L\) will be doubled.  The amplitude of
a gravitational wave is expressed as a dimensionless quantity giving the
change in distance per unit distance --- a shear strain in space.

Any apparatus capable of accurately measuring the distance between two
objects, or the change in the distance between two objects, as a function
of time is, in principle, capable of detecting a gravitational wave.  The
question of the effectiveness of the apparatus in accomplishing this task
is one of sensitivity and signal-to-noise ratio.  One design being actively
pursued by several research groups today is based on a Michelson-type laser
interferometer.  In this design, a laser beam is split and sent down the
two orthogonal arms of the apparatus, and then at the end of each is
reflected off a suspended mass back toward the splitter where the beams are
recombined \cite{abramovici:1992, caron1997, danzmann1997, luck1997,
kawabe1997}.  If the length of one or the other arms changes, the result
will be a change in the differential phase between the returning laser
beams and so a passing gravitational wave will appear as a shift in the
interference pattern at the output of the interferometer.  A variation on
this technique that is also being pursued by several research groups is a
detector consisting of an Earth-based transmitter/receiver and a spacecraft
carrying a radio or optical transponder.  There are currently proposals to
launch specially-designed vehicles dedicated to this task using optical
links from the ground to the vehicle and back \cite{ni2002}.  It is also
possible to use existing missions.  Deep-space vehicles generally carry
radio transponders that are used as aids to navigation.  By maintaining a
coherent up-link/down-link channel at high frequency stability with
appropriate modulations one can measure the range and range-rate to the
spacecraft with high precision and thereby search for the passage of a
gravitational wave through the solar system \cite{bertotti1998}.

\begin{figure}
\begin{center}
\begin{picture}(0,0)%
\includegraphics{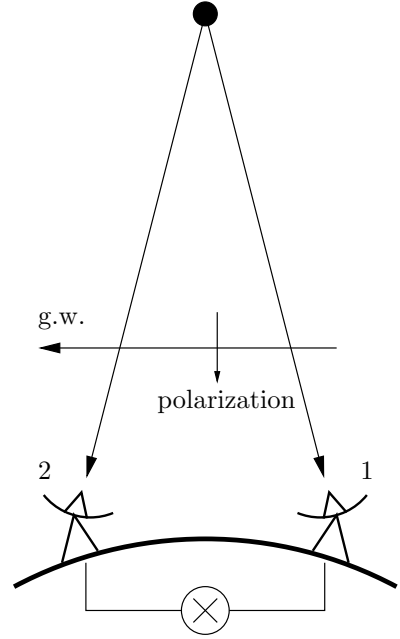}%
\end{picture}%
\setlength{\unitlength}{3947sp}%
\begingroup\makeatletter\ifx\SetFigFont\undefined%
\gdef\SetFigFont#1#2#3#4#5{%
  \reset@font\fontsize{#1}{#2pt}%
  \fontfamily{#3}\fontseries{#4}\fontshape{#5}%
  \selectfont}%
\fi\endgroup%
\begin{picture}(2446,3990)(-22,-3068)
\put(151,-2086){\makebox(0,0)[lb]{\smash{\SetFigFont{10}{12.0}{\familydefault}{\mddefault}{\updefault}2}}}
\put(2176,-2086){\makebox(0,0)[lb]{\smash{\SetFigFont{10}{12.0}{\familydefault}{\mddefault}{\updefault}1}}}
\put(901,-1636){\makebox(0,0)[lb]{\smash{\SetFigFont{10}{12.0}{\familydefault}{\mddefault}{\updefault}polarization}}}
\put(151,-1111){\makebox(0,0)[lb]{\smash{\SetFigFont{10}{12.0}{\familydefault}{\mddefault}{\updefault}g.w.}}}
\end{picture}
\end{center}
\caption{A gravitational wave moving across the lines of sight in the plane
of a two-station interferometer.  One of the gravitational wave's
polarization axes is aligned (ignoring parallax) with the lines of sight.}
\label{fig1}
\end{figure}

In this document, another type of gravitational wave detection apparatus
similar to this last technique is introduced and investigated.  Figure
\ref{fig1} shows a radio interferometer arrangement with a distant radio
source and two receiving elements.
It also shows a gravitational wave travelling across the lines of sight to
the radio source in the plane defined by the source and the two
interferometer elements (to be referred to, hereinafter, as the plane of
the interferometer).  For the purpose of illustration, one of the
gravitational wave's polarization axes is aligned (ignoring parallax) with
the lines of sight.  Because the gravitational wave travels at finite
speed, any perturbation to distances transverse to its direction of travel
will influence the signal arriving at receiver 1 in advance of its
influence on the signal arriving at receiver 2.  The presence of a
gravitational wave will, therefore, manifest itself as a time-varying
difference between the arrival times of radio signals at the receivers ---
a time-varying differential phase between the stations.  Such a detector is
similar in operation to the single arm, spacecraft-based, coherent
up-link/down-link type of detector.  In this case, however, a second arm is
introduced to act as a phase reference for the first which permits the use
of a broad-band radio signal.  What follows is an analysis of the magnitude
of the differential phase between the stations when a gravitational wave is
present, and a preliminary assessment of the signal-to-noise ratio in this
type of detector.

We will consider a radio signal emanating from the radio source toward each
of the receiving stations.  Each radio signal is treated as a null geodesic
in space-time, originating on the radio source's world line.  The
intersections of the geodesics with the world lines of the receivers define
events in space-time giving the co-ordinate times of arrival of the signal
at each receiving element.  By determining the difference in the times of
emission of the signals being received simultaneously at the two receiving
elements, and multiplying this time difference by the angular frequency of
the radio emission one obtains the differential phase between the two
receiving elements.  The calculation will first be done for arbitrary
receiver locations and arbitrary gravitational waveforms, then a simplified
interferometer arrangement will be introduced in order to more easily
discuss the response to specific gravitational waves.  Throughout the
calculations, the units for length will be chosen so as to make the
numerical value of the speed of light, \(c\), equal to 1.

\section{Null Geodesics in a Plane Gravitational Wave Space-Time}

For simplicity, we assume a Minkowski background for the geometry of
space-time.  Rather than solving the geodesic equation for fixed radio
source and receiver positions in an arbitrarily-oriented gravitational
wave, we shall choose to fix the orientation of the gravitational wave and
adjust the geodesic's end-points.  The co-ordinate system is defined as
follows:  the radio source locates the origin of the spatial co-ordinates,
the gravitational wave vector sets the negative-\(z\) direction and the two
polarization axes of the gravitational wave set the \(x\) and \(y\)
directions.  The origin of the time co-ordinate is arbitrary.  Space-time
geometry is described by the metric tensor
\begin{equation}
:g_\mu\nu:
   = :\eta_\mu\nu: + :h_\mu\nu:,
\end{equation}
where the background Minkowski metric tensor is
\begin{equation}
:\eta_\mu\nu:
   = \diag ( -1, 1, 1, 1 ),
\end{equation}
and the gravitational wave is described by the tensor \(:h_\mu\nu:\)
given by
\begin{equation}
:h_\mu\nu:
   = h( z + t ) :{e^{+}}_\mu\nu:.
\end{equation}
\(:{e^{+}}_\mu\nu:\) is the gravitational wave's polarization tensor which,
in this case, is
\begin{equation}
:{e^{+}}_\mu\nu:
   = \diag ( 0, 1, -1, 0 ),
\end{equation}
and \(h( z + t )\) is the gravitational wave's amplitude and describes a
waveform propagating at the speed of light in the negative \(z\) direction.
\(h( z + t )\) carries units of strain --- change in length per unit length
--- making it dimensionless.  Since the radio source defines the origin of
the spatial co-ordinates, the calculations are being done in its rest frame
so co-ordinate time corresponds to the proper time measured by a clock
carried by the radio source.  This may or may not agree with the proper
times measured at the receiving elements.

We assume the amplitude of the gravitational wave is small, \(\magnitude{h}
\ll 1\), and carry out calculations to first order in \(h\).  Altogether,
the line element describing the geometry of space-time is
\begin{multline}
\label{eqn4}
\diff s^{2}
   = -\diff t^{2} + \left[ 1 + h( z + t ) \right] \diff x^{2} + \\ \left[ 1
   - h( z + t ) \right] \diff y^{2} + \diff z^{2}.
\end{multline}
The only non-zero Christoffel symbols for this line element are
\begin{gather*}
:\Gamma^t_xx: = -:\Gamma^t_yy:
   = \frac{1}{2} h'( z + t )
   \\
:\Gamma^z_xx: = -:\Gamma^z_yy:
   = -\frac{1}{2} h'( z + t )
   \\
:\Gamma^x_tx: = :\Gamma^x_xt:
   = \frac{1}{2} h'( z + t )
   \\
:\Gamma^x_zx: = :\Gamma^x_xz:
   = \frac{1}{2} h'( z + t )
   \\
:\Gamma^y_ty: = :\Gamma^y_yt:
   = -\frac{1}{2} h'( z + t )
   \\
:\Gamma^y_zy: = :\Gamma^y_yz:
   = -\frac{1}{2} h'( z + t ),
\end{gather*}
where \(h'\) is the derivative of \(h\) with respect to its argument:
\(h'(z + t) = \frac{\diff}{\diff (z + t)} h( z + t )\).  A geodesic in the
space-time is a solution of the geodesic equation,
\begin{equation}
\frac{\diff^{2} :x^\mu:}{\diff \lambda^{2}} + :\Gamma^\mu_\alpha\beta:
\frac{\diff :x^\alpha:}{\diff \lambda} \frac{\diff :x^\beta:}{\diff
\lambda}
   = 0,
\end{equation}
and is described by four functions of an affine parameter:  \(x( \lambda
)\), \(y( \lambda )\), \(z( \lambda )\), and \(t( \lambda )\), where
\(\lambda\) parameterizes events along the the geodesic.  Using the
Christoffel symbols from above in the geodesic equation leads to the
following four coupled, second-order, non-linear, differential equations
for \(x( \lambda )\), \(y( \lambda )\), \(z( \lambda )\), and \(t( \lambda
)\) along a geodesic in this space-time:
\begin{subequations}
\label{eqn14}
\begin{gather}
\label{eqn1}
\frac{\diff^{2} t}{\diff \lambda^{2}} + \frac{1}{2} h'( z + t ) \left[
\left( \frac{\diff x}{\diff \lambda} \right)^{2} - \left( \frac{\diff
y}{\diff \lambda} \right)^{2} \right]
   = 0,
   \\
\label{eqn12}
\frac{\diff^{2} x}{\diff \lambda^{2}} + h'( z + t) \left[ \frac{\diff
z}{\diff \lambda} + \frac{\diff t}{\diff \lambda} \right] \frac{\diff
x}{\diff \lambda}
   = 0,
   \\
\label{eqn13}
\frac{\diff^{2} y}{\diff \lambda^{2}} - h'( z + t) \left[ \frac{\diff
z}{\diff \lambda} + \frac{\diff t}{\diff \lambda} \right] \frac{\diff
y}{\diff \lambda}
   = 0,
   \\
\label{eqn2}
\frac{\diff^{2} z}{\diff \lambda^{2}} - \frac{1}{2} h'( z + t ) \left[
\left( \frac{\diff x}{\diff \lambda} \right)^{2} - \left( \frac{\diff
y}{\diff \lambda} \right)^{2} \right]
   = 0.
\end{gather}
\end{subequations}
A unique solution to these equations requires eight boundary/initial
conditions, seven of which are provided by our knowledge of the end points
of the geodesics:  we know the spatial co-ordinates of the radio source and
the spatial co-ordinates of the receiving element and also either the time
of emission or the time of reception.  If we choose \(\lambda = 0\) to
correspond to the emission event and \(\lambda = 1\) the reception event on
the radio signal's null geodesic, then the boundary conditions are
\begin{subequations}
\begin{align}
t(\lambda = 0)
   & = t_{0},
   &
t(\lambda = 1)
   & = t_{i},
   \\
x(\lambda = 0)
   & = 0,
   &
x(\lambda = 1)
   & = x_{i},
   \\
y(\lambda = 0)
   & = 0,
   &
y(\lambda = 1)
   & = y_{i},
   \\
z(\lambda = 0)
   & = 0,
   &
z(\lambda = 1)
   & = z_{i},
\end{align}
\end{subequations}
where \(x_{i}\), \(y_{i}\), \(z_{i}\), and \(t_{i}\) are the space-time
co-ordinates of the reception event at the \(i^{\text{th}}\) receiving
element, and \(t_{0}\) is the co-ordinate time of emission of the radio
signal.  The subscripts on the reception event's co-ordinates will be used
later to enumerate them when multiple receiving elements are considered.
We have introduced both \(t_{0}\) and \(t_{i}\) but only one of the two is
known.  The other is to be found from the solution to the geodesic
equation.  The eighth constraint on the solution will be provided by
imposing the requirement that the geodesic be null.

Adding \eqref{eqn1} to \eqref{eqn2} gives
\begin{equation}
\label{eqn10}
\frac{\diff^{2}}{\diff \lambda^{2}} ( z + t )
   = 0
\end{equation}
which requires that whatever form \(z(\lambda)\) and \(t(\lambda)\) take,
their sum must be a linear function of \(\lambda\).  We introduce \(r_{i} =
\sqrt{x_{i}^{2} + y_{i}^{2} + z_{i}^{2}}\) to be the distance from the
radio source to the \(i^{\text{th}}\) receiving element in the absence of
the gravitational wave.  This provides the time-of-flight, \(t_{i} -
t_{0}\), for the radio signal from source to receiver in the absence of the
gravitational wave.  The presence of a gravitational wave alters this
time-of-flight to the new value \(a_{1} r_{i}\) where \(a_{1}\) is an
unknown dimensionless quantity to be found from the solution to the
geodesic equation.  The co-ordinate time of emission, \(t_{0}\), of the
radio signal can be expressed in terms of the the co-ordinate time of
reception and the time-of-flight in the presence of a gravitational wave as
\begin{equation}
\label{eqn3}
t_{0}
   = t_{i} - a_{1} r_{i}.
\end{equation}
Using this and the boundary conditions on \(z( \lambda )\) and \(t(
\lambda)\), we can write the solution to \eqref{eqn10} as
\begin{equation}
\label{eqn11}
z( \lambda ) + t( \lambda )
   = \begin{cases}
   ( z_{i} + a_{1} r_{i} ) \lambda + t_{0},
   \\
   ( z_{i} + a_{1} r_{i} ) \lambda + t_{i} - a_{1} r_{i},
   \end{cases}
\end{equation}
where it is expressed in terms of either \(t_{0}\) or \(t_{i}\) (depending
on which of the two is known), and the unknown \(a_{1}\) which appears here
as an integration constant.  Having the solution in \eqref{eqn11} succeeds
in decoupling each of the equations for \(x(\lambda)\) and \(y(\lambda)\)
in \eqref{eqn14} from the other three since it is only \(z + t\) and its
derivatives that appear in those two equations.

Introducing
\begin{equation}
\label{eqn15}
\tau( \lambda )
   = z( \lambda ) + t( \lambda ),
\end{equation}
we can write \(h'(z + t)\) as
\begin{equation*}
h'( z + t )
   = \frac{\diff}{\diff \tau} h( \tau )
   = \left( \frac{\diff \tau}{\diff \lambda} \right)^{-1}
   \frac{\diff}{\diff \lambda} h[ \tau( \lambda ) ].
\end{equation*}
From \eqref{eqn11}, \(\diff \tau / \diff \lambda\) is a constant
(independent of \(\lambda\)) equal to
\begin{equation}
\label{eqn16}
\frac{\diff \tau}{\diff \lambda}
   = \frac{\diff z}{\diff \lambda} + \frac{\diff t}{\diff \lambda}
   = z_{i} + a_{1} r_{i}.
\end{equation}
The \(x\) equation in \eqref{eqn12} then becomes
\begin{equation*}
\frac{\diff^{2} x}{\diff \lambda^{2}} + \frac{\diff x}{\diff \lambda}
\frac{\diff}{\diff \lambda} h( \tau )
   = 0
\end{equation*}
which can be integrated once to get
\begin{equation*}
\frac{\diff x}{\diff \lambda}
   = x_{i} [ 1 + a_{2} ] \ee^{-h( \tau )}
   \approx x_{i} [ 1 + a_{2} ] [ 1 - h( \tau ) ],
\end{equation*}
and then integrated again to get
\begin{subequations}
\begin{equation}
x( \lambda )
   = x_{i} \Bigl[ 1 + a_{2} \Bigr] \Bigl[ \lambda - \int_{0}^{\lambda} h[
   \tau(\lambda') ] \diff \lambda' + a_{3} \Bigr],
\end{equation}
where \(a_{2}\) and \(a_{3}\) are constants of integration.  The boundary
conditions on \(x(\lambda)\) require
\begin{gather}
a_{2}
   = \int_{0}^{1} h[ \tau(\lambda') ] \diff \lambda',
   \\
a_{3}
   = 0.
\end{gather}
\end{subequations}
Similarly, the \(y\) equation in \eqref{eqn13} becomes
\begin{equation*}
\frac{\diff^{2} y}{\diff \lambda^{2}} - \frac{\diff y}{\diff \lambda}
\frac{\diff}{\diff \lambda} h( \tau )
   = 0,
\end{equation*}
and by comparison with the \(x\) equation and its boundary conditions, it
can be verified that the solution for \(y( \lambda )\), satisfying its
boundary conditions, is
\begin{equation}
y( \lambda )
   = y_{i} \Bigl[ 1 - a_{2} \Bigr] \Bigl[ \lambda + \int_{0}^{\lambda} h[
   \tau(\lambda') ] \diff \lambda' - a_{3} \Bigr]
\end{equation}
where \(a_{2}\) and \(a_{3}\) are the same integration constants as in the
solution for \(x( \lambda )\).

For \(z(\lambda)\) in \eqref{eqn2}, we need the derivatives of
\(x(\lambda)\) and \(y(\lambda)\) but we only need them to
\(0^{\text{th}}\) order in \(h\) because higher order terms will result in
quadratic and higher order terms when multiplied by \(h'\) in the
differential equation.  These derivatives are
\begin{align*}
\frac{\diff x}{\diff \lambda}
   & \approx x_{i},
   &
\frac{\diff y}{\diff \lambda}
   & \approx y_{i}.
\end{align*}
With these, the \(z\) equation in \eqref{eqn2} becomes
\begin{equation*}
\frac{\diff^{2} z}{\diff \lambda^{2}} - \frac{1}{2} \left( \frac{x_{i}^{2}
- y_{i}^{2}}{\diff \tau / \diff \lambda} \right) \frac{\diff}{\diff
\lambda} h( \tau)
   = 0
\end{equation*}
which can be integrated twice to give
\begin{subequations}
\begin{equation}
z( \lambda )
   = z_{i} \Bigl[ \lambda + \frac{1}{2} \frac{x_{i}^{2} - y_{i}^{2}}{(
   \diff \tau / \diff \lambda ) z_{i}} \int_{0}^{\lambda} h[ \tau( \lambda'
   ) ] \diff \lambda' - a_{4} \lambda - a_{5} \Bigr]
\end{equation}
where \(a_{4}\) and \(a_{5}\) are integration constants.  The boundary
conditions on \(z(\lambda)\) require
\begin{gather}
a_{4}
   = \frac{1}{2} \frac{x_{i}^{2} - y_{i}^{2}}{( \diff \tau / \diff \lambda
   ) z_{i}} \int_{0}^{1} h[ \tau( \lambda') ] \diff \lambda',
   \\
a_{5}
   = 0.
\end{gather}
\end{subequations}

The eighth and final constraint on the solution is obtained from the
requirement that a space-time interval along the radio signal's geodesic be
null.  Setting \(\diff s^{2} = 0\) in \eqref{eqn4} yields
\begin{multline*}
\left( \frac{\diff t}{\diff \lambda} \right)^{2}
   = \left[ 1 + h( z + t ) \right] \left( \frac{\diff x}{\diff \lambda}
   \right)^{2} + \\ \left[ 1 - h( z + t) \right] \left( \frac{\diff
   y}{\diff \lambda} \right)^{2} + \left( \frac{\diff z}{\diff \lambda}
   \right)^{2},
\end{multline*}
and using \eqref{eqn15} for \(\diff t / \diff \lambda\), this becomes
\begin{multline}
\label{eqn9}
\left( \frac{\diff \tau}{\diff \lambda} \right)^{2} - 2 \frac{\diff
\tau}{\diff \lambda} \frac{\diff z}{\diff \lambda}
   = \\ [ 1 + h( \tau ) ] \left( \frac{\diff x}{\diff \lambda} \right)^{2}
   + [ 1 - h( \tau ) ] \left( \frac{\diff y}{\diff \lambda} \right)^{2}.
\end{multline}
Now, to first order in \(h\)
\begin{gather*}
\frac{\diff x}{\diff \lambda}
   = x_{i} \Bigl[ 1 - H( \tau ) \Bigr],
   \\
\frac{\diff y}{\diff \lambda}
   = y_{i} \Bigl[ 1 + H( \tau ) \Bigr],
   \\
\frac{\diff z}{\diff \lambda}
   = z_{i} \Bigl[ 1 + \frac{1}{2} \frac{x_{i}^{2} - y_{i}^{2}}{( \diff \tau
   / \diff \lambda ) z_{i}} H( \tau ) \Bigr],
\end{gather*}
where
\begin{equation*}
H( \tau )
   = h( \tau ) - \int_{0}^{1} h[ \tau( \lambda' ) ] \diff \lambda'.
\end{equation*}
With these, the constraint in \eqref{eqn9} becomes
\begin{multline*}
\left( \frac{\diff \tau}{\diff \lambda} \right)^{2} - 2 \frac{\diff
\tau}{\diff \lambda} z_{i} - ( x_{i}^{2} - y_{i}^{2} ) H
   = \\ x_{i}^{2} [ 1 + h( \tau ) - 2 H ] + y_{i}^{2} [ 1 - h( \tau ) + 2 H
   ],
\end{multline*}
or
\begin{equation*}
\frac{1}{r_{i}^{2}} \left( \frac{\diff \tau}{\diff \lambda} - z_{i}
\right)^{2}
   = 1 + \frac{x_{i}^{2} - y_{i}^{2}}{r_{i}^{2}} \int_{0}^{1} h[ \tau(
   \lambda' ) ] \diff \lambda'.
\end{equation*}
Using \eqref{eqn16} for \(\diff \tau / \diff \lambda\) and recalling that
there are two expressions for \(\tau\) depending on which end-point of the
geodesic is known, this becomes
\begin{align*}
a_{1}^{2}
   & = 1 + \frac{x_{i}^{2} - y_{i}^{2}}{r_{i}^{2}} \int_{0}^{1} h[ \tau(
   \lambda' ) ] \diff \lambda'
   \\
   & = \begin{cases}
   1 + \frac{x_{i}^{2} - y_{i}^{2}}{r_{i}^{2}} \int_{0}^{1} h[ ( z_{i} +
   a_{1} r_{i} ) \lambda' + t_{0}  ] \diff \lambda',
   \\
   1 + \frac{x_{i}^{2} - y_{i}^{2}}{r_{i}^{2}} \int_{0}^{1} h[ ( z_{i} +
   a_{1} r_{i} ) \lambda' + t_{i} - a_{1} r_{i}  ] \diff \lambda'.
   \end{cases}
\end{align*}
This constitutes a transcendental equation for the integration constant
\(a_{1}\) --- the last unknown in the solution for the geodesic.  This
equation may be solved by recognizing that the right-hand side is equal to
1 plus a quantity proportional to \(h\), and so requiring the left-hand
side to also be equal to 1 plus a quantity proportional to \(h\).
Consequently it is convenient to write
\begin{subequations}
\begin{equation}
a_{1}
   = 1 + \frac{1}{2} a_{6}
\end{equation}
where \(a_{6}\) must be proportional to \(h\).  Making this substitution
and discarding terms higher than first order in \(h\) yields\footnote{The
\(a_{6}\)'s inside the integrand on the right-hand side are discarded.  To
do this carefully, one expands the integrand in a Taylor series in powers
of \(a_{6}\).  Since \(a_{6}\) must be proportional to \(h\), only the
zeroth-order term in that series need be retained as all higher order terms
are quadratic or higher in \(h\).}
\begin{equation*}
a_{6}
   = \begin{cases}
   \frac{x_{i}^{2} - y_{i}^{2}}{r_{i}^{2}} \int_{0}^{1} h[ ( z_{i} +
   r_{i} ) \lambda' + t_{0} ] \diff \lambda',
   \\
   \frac{x_{i}^{2} - y_{i}^{2}}{r_{i}^{2}} \int_{0}^{1} h[ ( z_{i} +
   r_{i} ) \lambda' + t_{i} - r_{i} ] \diff \lambda'.
   \end{cases}
\end{equation*}
A change of variables in the integrals from \(\lambda'\) to \(\tau\) turns
these expressions for \(a_{6}\) into
\begin{equation}
a_{6}
   = \begin{cases}
   \frac{x_{i}^{2} - y_{i}^{2}}{( z_{i} + r_{i} ) r_{i}^{2}}
   \int_{t_{0}}^{t_{0} + z_{i} + r_{i}} h( \tau ) \diff \tau,
   \\
   \frac{x_{i}^{2} - y_{i}^{2}}{( z_{i} + r_{i} ) r_{i}^{2}} \int_{t_{i} -
   r_{i}}^{t_{i} + z_{i}} h( \tau ) \diff \tau.
   \end{cases}
\end{equation}
\end{subequations}

Finally, using the expressions for \(a_{6}\) in \(a_{1}\) and substituting
the results into \eqref{eqn3} we obtain
\begin{align}
\label{eqn19}
t_{i}
   & = t_{0} + r_{i} + \frac{x_{i}^{2} - y_{i}^{2}}{2 ( z_{i} + r_{i} )
   r_{i}} \int_{t_{0}}^{t_{0} + z_{i} + r_{i}} h( \tau ) \diff \tau,
   \\\intertext{and}
\label{eqn17}
t_{0}
   & = t_{i} - r_{i} - \frac{x_{i}^{2} - y_{i}^{2}}{2 ( z_{i} + r_{i} )
   r_{i}} \int_{t_{i} - r_{i}}^{t_{i} + z_{i}} h( \tau ) \diff \tau.
\end{align}
These give, respectively, the co-ordinate time of reception as a function
of the receiver position and the co-ordinate time of emission,
\eqref{eqn19}, and the co-ordinate time of emission as a function of the
receiver position and the co-ordinate time of reception, \eqref{eqn17}.
From these we see that the time-of-flight of the radio signal is given by
\(r_{i}\) plus a term expressing the perturbation of the radio wave's
travel time due to its passage through the gravitational wave en route to
the point of reception.

\section{The Differential Phase in a Two-Element Radio Interferometer}

Consider an interferometer system consisting of two receiving elements and
a data acquisition system at each element which is capable of establishing
co-ordinate time and recording the radio signal being received as a
function of this parameter.  In general, the signals being received at the
same co-ordinate time at each element, \(t_{1} = t_{2} = t\), will have
been emitted from the radio source at different co-ordinate times.  The
difference in the times of emission can be obtained from \eqref{eqn17} and
is
\begin{multline*}
\Delta t_{0}
   = r_{2} - r_{1} + \\ \frac{1}{2} \Biggl[ \frac{x_{2}^{2} - y_{2}^{2}}{(
   z_{2} + r_{2} ) r_{2}} \int_{t - r_{2}}^{t + z_{2}} h( \tau ) \diff \tau
   - \\ \frac{x_{1}^{2} - y_{1}^{2}}{( z_{1} + r_{1} ) r_{1}} \int_{t -
   r_{1}}^{t + z_{1}} h( \tau ) \diff \tau \Biggr].
\end{multline*}
\(r_{2} - r_{1}\) is the flat-space geometric delay for the interferometer
and the remainder of the expression represents an additional contribution
to the differential emission time arising from the presence of a
gravitational wave.  If the radio receiving elements are tuned to an
angular frequency of \(\omegar\) (in the rest frame of the radio source),
then this differential emission time translates into a differential phase
of the received signals given by \(\Delta \phi = \omegar \Delta t_{0}\), or
\begin{multline}
\label{eqn5}
\Delta \phi( t )
   = \omegar ( r_{2} - r_{1} ) + \\ \frac{\omegar}{2} \Biggl[
   \frac{x_{2}^{2} - y_{2}^{2}}{( z_{2} + r_{2} ) r_{2}} \int_{t -
   r_{2}}^{t + z_{2}} h( \tau ) \diff \tau - \\ \frac{x_{1}^{2} -
   y_{1}^{2}}{( z_{1} + r_{1}) r_{1}} \int_{t - r_{1}}^{t + z_{1}} h( \tau
   ) \diff \tau \Biggr].
\end{multline}

The differential phase in \eqref{eqn5} is quite general.  In principle, for
example, the expression continues to hold for radio receiving elements that
are in motion with respect to the radio source and will correctly account
for effects such as the Doppler shifting of the radio signals.  The
expression, however, is arrived at in the rest frame of the radio source.
This means, for example, that the co-ordinate time, \(t_{i}\), at each
receiving element is the proper time measured by a clock carried by the
radio source, \emph{not} by a clock carried by that receiving element.  In
practise it is local proper time that is available at each receiving
element.  The implementation of this gravitational wave observation
technique will require knowledge of the transformation from local proper
time to co-ordinate time.  For a discussion of this transformation, see
\cite{hellings1986}.

\begin{figure}
\begin{center}
\begin{picture}(0,0)%
\includegraphics{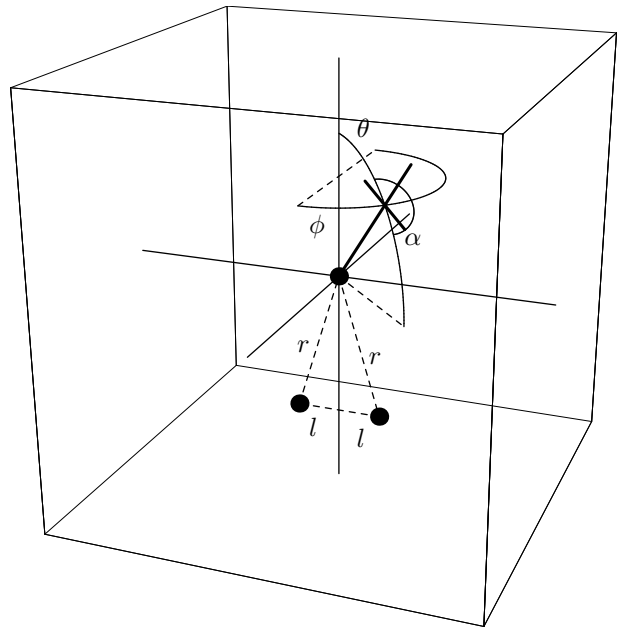}%
\end{picture}%
\setlength{\unitlength}{3947sp}%
\begingroup\makeatletter\ifx\SetFigFont\undefined%
\gdef\SetFigFont#1#2#3#4#5{%
  \reset@font\fontsize{#1}{#2pt}%
  \fontfamily{#3}\fontseries{#4}\fontshape{#5}%
  \selectfont}%
\fi\endgroup%
\begin{picture}(3817,3900)(1,-3061)
\put(1876,-1861){\makebox(0,0)[lb]{\smash{\SetFigFont{10}{12.0}{\familydefault}{\mddefault}{\updefault}\(l\)}}}
\put(2176,-1936){\makebox(0,0)[lb]{\smash{\SetFigFont{10}{12.0}{\familydefault}{\mddefault}{\updefault}\(l\)}}}
\put(1876,-586){\makebox(0,0)[lb]{\smash{\SetFigFont{10}{12.0}{\familydefault}{\mddefault}{\updefault}\(\phi\)}}}
\put(2476,-661){\makebox(0,0)[lb]{\smash{\SetFigFont{10}{12.0}{\familydefault}{\mddefault}{\updefault}\(\alpha\)}}}
\put(1801,-1336){\makebox(0,0)[lb]{\smash{\SetFigFont{10}{12.0}{\familydefault}{\mddefault}{\updefault}\(r\)}}}
\put(2251,-1411){\makebox(0,0)[lb]{\smash{\SetFigFont{10}{12.0}{\familydefault}{\mddefault}{\updefault}\(r\)}}}
\put(2176, 14){\makebox(0,0)[lb]{\smash{\SetFigFont{10}{12.0}{\familydefault}{\mddefault}{\updefault}\(\theta\)}}}
\end{picture}
\end{center}
\caption{The orientation of the interferometer with respect to the
gravitational wave's wave-vector and polarization.}
\label{fig5}
\end{figure}

We now consider a special case for the interferometer configuration
consisting of two receivers at rest with respect to and equally distant
from the radio source, \(r_{1} = r_{2} = r\), and separated from each other
by a distance \(2 l\) where \(l / r \ll 1\).  The geometry of the
arrangement is depicted in Figure \ref{fig5}.
We introduce three angles to describe the interferometer's orientation with
respect to the gravitational wave:  two to describe the direction from
which the gravitational wave is approaching and a third to describe the
orientation of its polarization axes.  The direction from which the
gravitational wave is approaching is defined by \((\theta, \phi)\), the
polar angle and azimuthal angle of a spherical polar co-ordinate system.
The polar axis, \(\theta = 0\), is directed toward the radio source,
bisecting the interferometer's baseline.  Also bisecting the
interferometer's baseline and perpendicular to the polar axis is the origin
of the \(\phi\) co-ordinate.  The orientation of the gravitational wave's
polarization tensor is given by \(\alpha\) which is measured in the plane
orthogonal to the gravitational wave vector at the point \((\theta,
\phi)\), positively from the meridian of constant \(\phi\) to the
polarization axis as shown in Figure \ref{fig5}.  The choice of which
polarization axis is used to define the angle \(\alpha\) is arbitrary since
the two possible choices simply differ from one another by a rotation of
\(\unit{\pi}{\rad}\) in the detector's output.

In the co-ordinate system used to obtain the differential phase in
\eqref{eqn5}, the receiver positions are given by
\begin{widetext}
\begin{multline*}
\begin{pmatrix}
\cos \alpha & \sin \alpha & 0 \\
-\sin \alpha & \cos \alpha & 0 \\
0 & 0 & 1
\end{pmatrix} \cdot \begin{pmatrix}
\cos \theta & 0 & -\sin \theta \\
0 & 1 & 0 \\
\sin \theta & 0 & \cos \theta
\end{pmatrix} \cdot \begin{pmatrix}
\cos \phi & \sin \phi & 0 \\
-\sin \phi & \cos \phi & 0 \\
0 & 0 & 1
\end{pmatrix} \cdot \begin{pmatrix}
0 \\ \pm l \\ -r
\end{pmatrix}
   = \\ r \begin{pmatrix}
   \sin \theta \cos \alpha \pm \frac{l}{r} ( \sin \phi \cos \theta \cos
   \alpha + \cos \phi \sin \alpha ) \\
   -\sin \theta \sin \alpha \mp \frac{l}{r} ( \sin \phi \cos \theta \sin
   \alpha - \cos \phi \cos \alpha ) \\
   -\cos \theta \pm \frac{l}{r} \sin \phi \sin \theta
   \end{pmatrix}.
\end{multline*}
\end{widetext}
The rotation matrices are those that will rotate the gravitational wave
vector in Figure \ref{fig5} back to the \(z\) axis while orienting its
polarization directions along the \(x\) and \(y\) axes.  From this,
\begin{equation*}
x^{2} - y^{2}
   = r^{2} \sin^{2} \theta \cos 2 \alpha \Bigl( 1 \pm 2 \frac{l}{r}
   \frac{\cos \theta \sin \phi}{\sin \theta \cos 2 \alpha} \Bigr)
\end{equation*}
and
\begin{equation*}
z + r
   = r ( 1 - \cos \theta ) \Bigl( 1 \pm \frac{l}{r} \frac{\sin \phi \sin
   \theta}{1 - \cos \theta} \Bigr),
\end{equation*}
so
\begin{multline*}
\frac{x^{2} - y^{2}}{(z + r) r}
   = ( 1 + \cos \theta ) \cos 2 \alpha \,\cdot \\ \Bigl[ 1 \pm \frac{l}{r}
   \Bigl( 2 \frac{\cos \theta \sin \phi}{\sin \theta \cos 2 \alpha} -
   \frac{\sin \phi \sin \theta}{1 - \cos \theta} \Bigr) \Bigr].
\end{multline*}
Using this expression in \eqref{eqn5} and considering terms in \(h
\frac{l}{r}\) to be second order and negligible, the interferometer's
differential phase reduces to
\begin{equation*}
\Delta \phi( t )
   = \frac{\omegar}{2} ( 1 + \cos \theta ) \cos 2 \alpha \int_{t - r \cos
   \theta - l \sin \theta \sin \phi}^{t - r \cos \theta + l \sin \theta
   \sin \phi} h( \tau ) \diff \tau.
\end{equation*}
Redefining the origin of the time co-ordinate so that \(t - r \cos \theta
\rightarrow t\), the differential phase for the simplified interferometer
configuration is finally
\begin{equation}
\label{eqn18}
\Delta \phi( t )
   = \frac{\omegar}{2} ( 1 + \cos \theta ) \cos 2 \alpha \int_{t - l \sin
   \theta \sin \phi}^{t + l \sin \theta \sin \phi} h( \tau ) \diff \tau.
\end{equation}

We can also express the differential phase in terms of the gravitational
wave's frequency-domain representation.  Introducing the Fourier transform
of the gravitational wave,
\begin{align}
h( \tau )
   & = \int_{-\infty}^{+\infty} \tilde{h}( \omega ) \ee^{\aye \omega \tau}
   \diff \omega,
   \\
\tilde{h}( \omega )
   & = \frac{1}{2 \pi} \int_{-\infty}^{+\infty} h( \tau ) \ee^{-\aye \omega
   \tau} \diff \tau,
\end{align}
the differential phase can be written as
\begin{multline}
\label{eqn8}
\Delta \phi( t )
   = ( 1 + \cos \theta ) \cos 2 \alpha \,\cdot \\ \int_{-\infty}^{+\infty}
   \frac{\omegar}{\omega} \tilde{h}( \omega ) \sin ( \omega l \sin \theta
   \sin \phi ) \ee^{\aye \omega t} \diff \omega.
\end{multline}

The output of the detector, \(\Delta \phi( t )\), does not mirror the
gravitational waveform, \(h( t )\).  Rather, the output is proportional to
the convolution of the waveform with a square time-domain window whose
amplitude and duration are determined by the orientation and baseline
length of the interferometer.  An alternative approach to the extraction of
the gravitational wave signal from the interferometer's output is to
differentiate the differential phase with respect to co-ordinate time.
Adjusting the origin of the time co-ordinate in \eqref{eqn18}, and
differentiating with respect to \(t\), one finds
\begin{equation*}
\frac{\diff \Delta \phi}{\diff t}
   = \frac{\omegar}{2} ( 1 + \cos \theta ) \cos 2 \alpha \bigl[ h( t ) - h(
   t - 2 l \sin \theta \sin \phi ) \bigr].
\end{equation*}
Although this equation is in the more standard form for the description of
the output of interferometric beam detectors \cite[equation
(5)]{schutz1999}, this method of signal extraction makes noise estimation
more complex and so this paper will confine itself to an examination of the
phase difference between the receiving stations alone.

\section{The Response to Specific Gravitational Waveforms}

\subsection{Monochromatic Gravitational Wave}

Here we shall consider the special case of a monochromatic gravitational
wave with angular frequency \(\omegagw\), described in the frequency domain
by
\begin{equation}
\tilde{h}( \omega )
   = h \frac{1}{2} \bigl[ \delta( \omega - \omegagw ) + \delta( \omega +
   \omegagw ) \bigr].
\end{equation}
In the time domain this is equivalent to \(h( t ) = h \cos \omegagw t\).
Substituting \(\tilde{h}( \omega )\) into \eqref{eqn8} results in a
time-averaged RMS differential phase of
\begin{equation*}
\Delta \rms{\phi}
   = \frac{h}{\sqrt{2}} \frac{\omegar}{\omegagw} \magnitude{( 1 + \cos
   \theta ) \cos 2 \alpha \sin ( \omegagw l \sin \theta \sin \phi )},
\end{equation*}
or
\begin{subequations}
\begin{equation}
\label{eqn7}
\Delta \rms{\phi}
   = \frac{1}{\sqrt{2}} \Phieff h
\end{equation}
where
\begin{equation}
\Phieff
   = \frac{\omegar}{\omegagw} \magnitude{( 1 + \cos \theta ) \cos 2 \alpha
   \sin ( \omegagw l \sin \theta \sin \phi )}.
\end{equation}
\end{subequations}
\(\Phieff\) gives the differential phase per unit of gravitational wave
strain and is a measure of the interferometer's sensitivity to
gravitational waves.  One can visualize \(\Phieff\) as follows.  Consider
replacing the interferometer with a simple distance measuring device
consisting of two test masses and a (very precise) measuring stick marked
off in radians of phase of the radio wavelength being observed by the
interferometer.  A passing gravitational wave will move the test masses
back and forth, and using the measuring stick one can read off the change
in the distance between them in radians.  \(\Phieff\) gives the number of
radians apart the test masses would need to be placed in order to observe
the same change in their separation, in radians, as there is differential
phase seen in the radio interferometer.  \(\Phieff\) shall be referred to
as the interferometer's ``effective phase length.''  The interferometer can
be assigned a ``natural'' phase length of \(\omegar 2 l\) (the separation
of the receiving stations in radians of the radio wavelength being
observed), and then the ratio
\begin{equation}
\efficiency
   = \frac{\Phieff}{\omegar 2 l},
\end{equation}
which ranges from 0 to \(\efficiency_{\text{max}} \approx 0.65\), can be
thought of as the detector's efficiency.

\begin{figure*}
\begin{center}
(a) \includegraphics{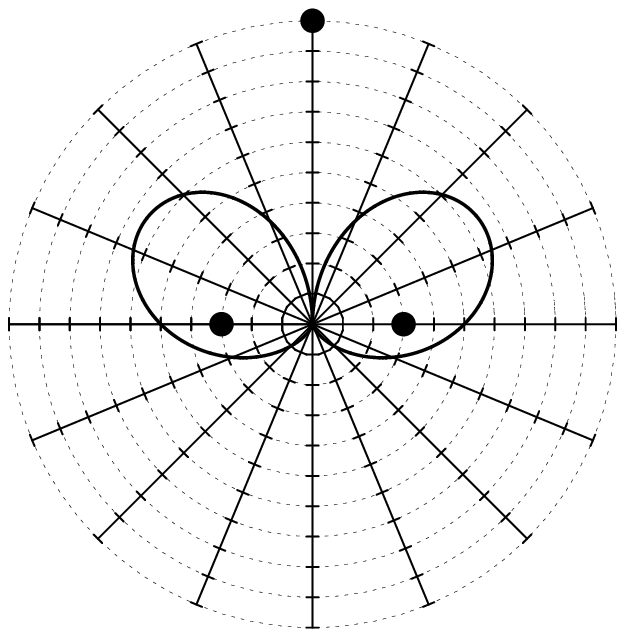} \includegraphics{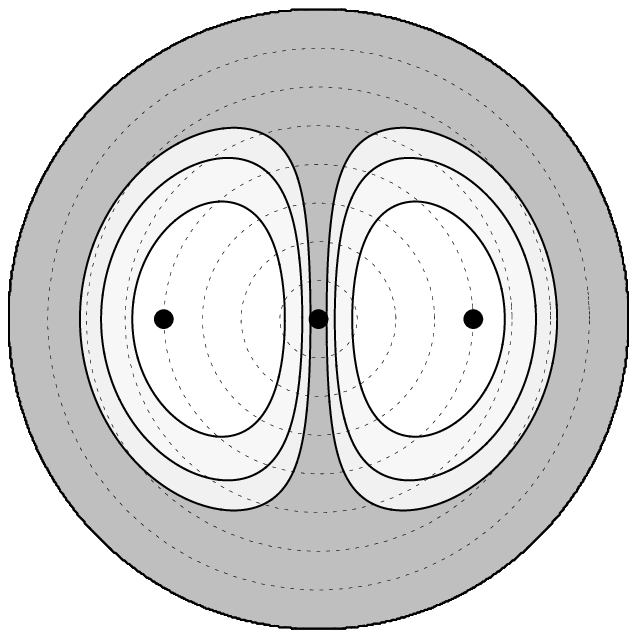}\\
(b) \includegraphics{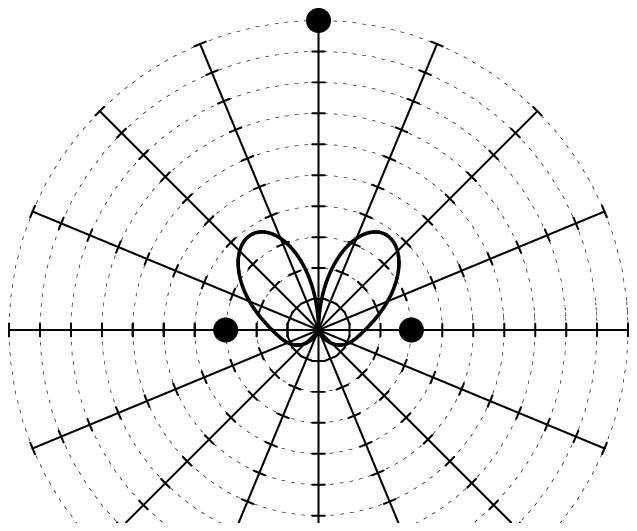} \includegraphics{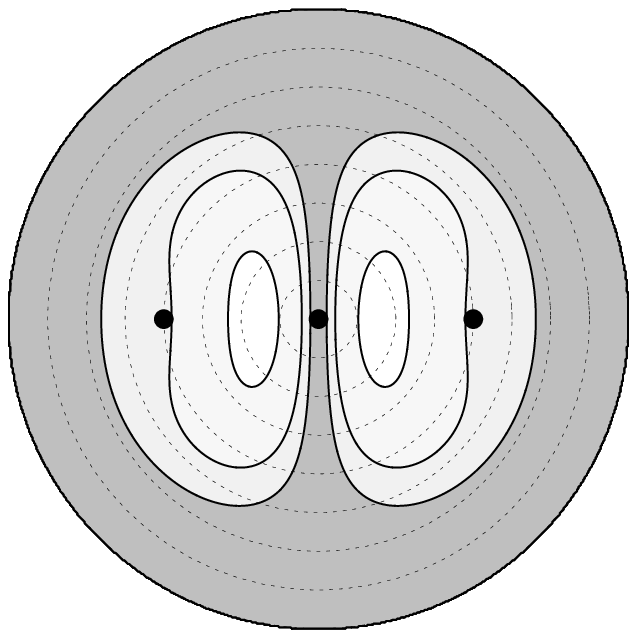}\\
(c) \includegraphics{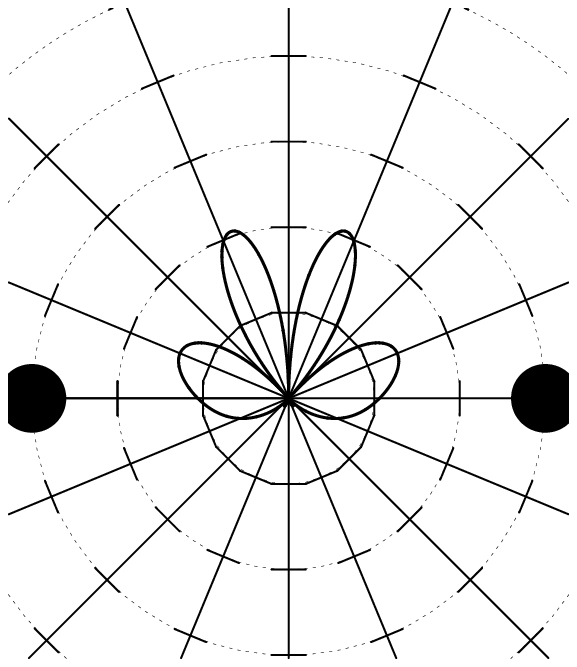} \includegraphics{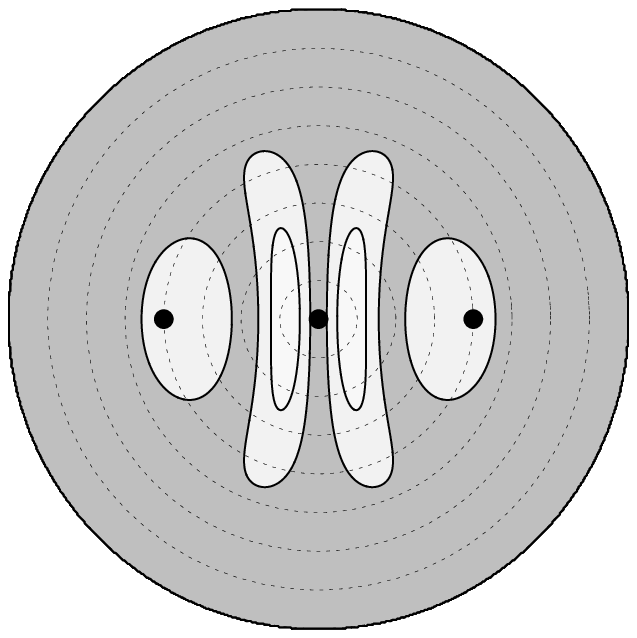}
\end{center}
\caption{The efficiency of the interferometer's response to monochromatic
gravitational waves at several wavelengths.  From top:  $\omegagw 2 l =
\unit{0}{\rad}$, $\unit{3 \pi / 2}{\rad}$, and $\unit{3 \pi}{\rad}$.  The
three filled circles in each plot indicate the orientation of the source
and receiving elements.  On the left are plots of the efficiency in the
plane of the interferometer with the outer ring being an efficiency of 1
and with rings drawn at intervals of 0.1.  On the right are the same
functions shown as contour plots over the full sky with contours drawn at
$\unit{-3}{\deci\Bel}$, $\unit{-6}{\deci\Bel}$, and $\unit{-9}{\deci\Bel}$
relative to the detector's best possible efficiency of
$\efficiency_{\text{max}} \approx 0.65$ (visible as the peak efficiency in
the top-leftmost plot).}
\label{fig4}
\end{figure*}

Since the efficiency is equivalent to the effective phase length normalized
with respect to a combination of interferometer parameters, it is not
useful for comparing the behaviour of different interferometers.  Since the
efficiency is, in fact, only a function of the orientation of the
gravitational wave with respect to the interferometer and the wavelength of
the gravitational wave in units of the separation of the two receiving
stations, it is a convenient way of discussing the dependence of any one
interferometer's sensitivity to the direction on the sky from which the
gravitational wave is approaching, the orientation of its polarization, and
its wavelength.  Plots of the efficiency showing its dependence on
direction and wavelength are shown in Figure \ref{fig4}.
These show, specifically, the behaviour of the function
\begin{equation*}
\efficiency
   = \frac{1}{2 \omegagw l} \magnitude{ ( 1 + \cos \theta ) \cos 2 \alpha
   \sin ( \omegagw l \sin \theta \sin \phi )}
\end{equation*}
both in the plane of the interferometer, \(\phi = \pm \pi / 2\) (figures on
the left), and over the full sky (figures on the right), for three choices
of gravitational wave wavelength:  from top to bottom \(\omegagw 2 l =
\unit{0}{\rad}\), \(\unit{3 \pi / 2}{\rad}\), and \(\unit{3 \pi}{\rad}\).
In all plots, the polarization of the gravitational wave is chosen so as to
maximize the efficiency (\(\alpha = \unit{0}{\rad}\)).  For the plots of
the efficiency in the plane of the interferometer, the scale is such that
the outer ring indicates an efficiency of 1 and rings are drawn at
intervals of 0.1.  The three filled circles indicate the orientation of the
interferometer elements with the uppermost circle indicating the direction
toward the radio source (\(\theta = \unit{0}{\rad}\)).  In the full-sky
contour plots, the filled circle at the centre of each plot indicates the
direction toward the radio source, while the outer ring corresponds to the
direction away from the radio source and the two filled circles at
\(\unit{\pi/2}{\rad}\) from the centre indicate the directions toward the
receiving elements.  Contours are drawn at \(\unit{-3}{\deci\Bel}\),
\(\unit{-6}{\deci\Bel}\), and \(\unit{-9}{\deci\Bel}\) relative to the
detectors best possible efficiency of \(\efficiency_{\text{max}} \approx
0.65\).  Several features are noteworthy:
\begin{itemize}
\item The direction of greatest sensitivity to gravitational waves is not
perpendicular to the interferometer's lines of sight to the radio source,
where one might expect the wave's polarization to be most favourable, but
rather lies in a direction displaced from the perpendicular toward the
radio source.

\item The interferometer is almost completely insensitive to gravitational
waves approaching from directions away from the radio source.

\item Compared to the response to longer wavelengths (Figure \ref{fig4}a),
the interferometer responds poorly to gravitational waves with wavelengths
shorter than the interferometer's baseline (Figure \ref{fig4}c).
\end{itemize}

\begin{figure}
\begin{center}
\begin{picture}(0,0)
\put(-10,50){\rotatebox{90}{\(\unit{\Phieff / \omegar}{(\second)}\)}}
\put(90,-5){\(\unit{\omegagw}{(\radian / \second)}\)}
\end{picture}%
\includegraphics{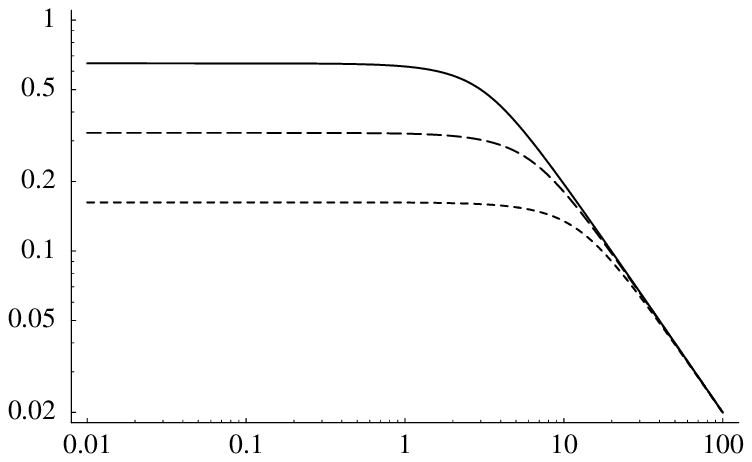}
\end{center}
\caption{The interferometer's effective length (in the direction of
greatest sensitivity) in units of light travel time as it depends on the
angular frequency of the gravitational wave and the length of the
interferometer's baseline.  For the purpose of illustration, baseline
lengths of $2l / c = \unit{1}{\second}$ (solid curve),
$\unit{\frac{1}{2}}{\second}$ (long dashes) and
$\unit{\frac{1}{4}}{\second}$ (short dashes) were chosen.  A larger
effective length means the generation of a larger differential phase for
the same gravitational wave amplitude and thus a more sensitive detector.}
\label{fig6}
\end{figure}

To more clearly see how the frequency response of the interferometer
depends on the baseline length, Figure \ref{fig6} shows the
interferometer's effective length in the direction of greatest sensitivity
as a function of the gravitational wave's angular frequency for three
choices of baseline.
For the purpose of illustration, some fairly long baseline lengths were
chosen:  in terms of the light crossing time, they have lengths of
\(\unit{1}{\second}\), \(\unit{\frac{1}{2}}{\second}\), and
\(\unit{\frac{1}{4}}{\second}\).  The gravitational waves with these
wavelengths have angular frequencies of \(\unit{2 \pi}{\radian /
\second}\), \(\unit{4 \pi}{\radian / \second}\) and \(\unit{8 \pi}{\radian
/ \second}\) respectively.  Recalling that the effective length is a
measure of the sensitivity of the interferometer to gravitational waves,
Figure \ref{fig6} shows a flat response to gravitational waves with
wavelengths longer than the interferometer's baseline.  The poor short
wavelength performance is attributable to the differential phase being the
convolution of the gravitational waveform with a square window whose
duration is comparable to the light-crossing time of the interferometer's
baseline:  gravitational waves with periods less than this are smoothed
out, diminishing their influence on the interferometer's output.  Lest one
think the solution to the poor high frequency performance is to use shorter
baselines, we also see from Figure \ref{fig6} that going to shorter
baselines does not improve the high frequency response at all.  It does
flatten the response to higher frequencies but it accomplishes this by
reducing the low frequency sensitivity, making it as poor as that at high
frequencies.  The conclusion is that longer baselines are better:  they
improve the low frequency response without, in fact, harming the high
frequency response.  The interferometer's flat response at low frequencies
can be seen analytically by checking that
\begin{equation*}
\lim_{\omegagw l \rightarrow 0} \frac{\diff \efficiency}{\diff \omegagw}
   = 0,
\end{equation*}
and the actual efficiency the interferometer asymptotes to at low
frequencies is given by
\begin{gather*}
\lim_{\omegagw l \rightarrow 0} \efficiency
   = \frac{1}{2} \magnitude{( 1 + \cos \theta ) \sin \theta \sin \phi \cos
   2 \alpha}.
\end{gather*}

It is interesting that although longer baselines do improve the overall
sensitivity of the interferometer, contrary to what one might expect upon
consideration of Figure \ref{fig1} it is not \emph{necessary} to construct
an interferometer with a long baseline in order to avoid being insensitive
to long wavelength gravitational waves.  One might have thought, for
example, that a gravitational wave with a wavelength much longer than the
separation of the receiving elements would have the two lines of sight
always sitting at approximately the same phase within the wave, thus making
it difficult to detect its presence.  However, the long wavelength also
means the gravitational wave's effect on the radio signal is integrated
over a longer segment of the line of sight.  The two effects balance each
other giving the interferometer a flat response at low frequencies.

Although this section has looked specifically at an exactly monochromatic
gravitational wave, the results can be reasonably applied to many
anticipated astrophysical sources of gravitational waves.  From
\eqref{eqn18}, the relationship between the differential phase and the
gravitational waveform only involves the integration of the waveform over
an interval roughly equivalent to the light-crossing time of the
interferometer's baseline.  The results of this section continue to hold so
long as the gravitational wave is approximately monochromatic over this
interval.  For radio receiving elements restricted to the surface of the
Earth, this is an interval less than \(\unit{40}{\milli\second}\).  Even
for space-based antennas separated by millions of kilometres the interval
is on the order of \(\unit{10}{\second}\).  Over these time scales, many
anticipated astrophysical sources of gravitational waves can be
approximated as exactly periodic.

\subsection{Black Hole Ring-Down Gravitational Wave}

One example of an astrophysical system that can't reasonably be
approximated as exactly periodic over tens of milliseconds is a perturbed
black hole.  When a spinning black hole is distorted, the irregularities of
its event horizon are radiated away in the form of gravitational waves.
The particular waveform generated by this process is called a ``black hole
ring-down'' waveform and is of the form \cite{costa2003, echeverria1989}
\begin{equation}
h( t )
   = \begin{cases}
   0, & t < \unit{0}{\second},
   \\
   h \ee^{-\omega_{\text{s}} t / (2 Q)} \sin \omega_{\text{s}} t, & t \ge
   \unit{0}{\second},
   \end{cases}
\end{equation}
where
\begin{equation}
\omega_{\text{s}}
   \approx ( \unit{200}{\kilo\rad/\sec} ) \left[ 1 - 0.63 ( 1 - a
   )^{\frac{3}{10}} \right] \left( \frac{M}{\sol{M}} \right)^{-1}
\end{equation}
is the centre frequency of the black hole's fundamental quadrupolar mode,
\begin{equation}
Q
   \approx 2 ( 1 - a )^{-\frac{9}{20}}
\end{equation}
is the ``quality factor,'' \(M / \sol{M}\) is the black hole's mass in
solar mass units, and \(a\) is the black hole's dimensionless angular
momentum (it's physical angular momentum is \(a M G / c^{2}\)).  From
\eqref{eqn18}, the interferometer's differential phase, \(\Delta \phi(
t)\), is related to the gravitational waveform by integration over an
interval of time.  Adjusting the origin of the time co-ordinate in
\eqref{eqn18} so that this interval just encounters the start of the
ring-down waveform at \(t = 0\), the differential phase produced by this
type of wave is
\begin{multline*}
\Delta \phi( t )
   = h \frac{\omegar}{2} ( 1 + \cos \theta ) \cos 2 \alpha \,\cdot \\
   \int_{t_{\text{start}}}^{t} \ee^{-\omega_{\text{s}} \tau / (2 Q)} \sin
   \omega_{\text{s}} \tau \diff \tau
\end{multline*}
for \(t \ge 0\), where
\begin{equation*}
t_{\text{start}}
   = \text{max}( 0, t - 2 l \sin \theta \sin \phi ).
\end{equation*}
This evaluates to
\begin{multline}
\Delta \phi( t )
   = h \frac{\omegar}{\omega_{\text{s}}} \frac{Q}{1 + 4 Q^{2}} ( 1 + \cos
   \theta ) \cos 2 \alpha \,\cdot \\ \left[ -\ee^{-\omega_{\text{s}} \tau /
   (2 Q)} ( 2 Q \cos \omega_{\text{s}} \tau + \sin \omega_{s} \tau )
   \right]_{t_{\text{start}}}^{t}.
\end{multline}

\begin{figure}
\begin{center}
\begin{picture}(0,0)
\put(-10,50){\rotatebox{90}{\(h( t ) / h\)}}
\put(100,-5){time (\(\milli \second\))} 
\end{picture}%
\includegraphics{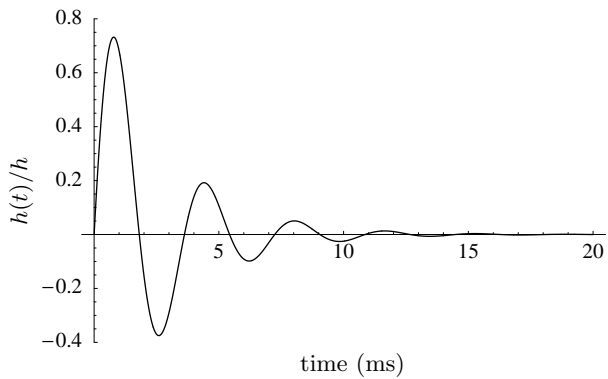}
\end{center}
\caption{The ring-down waveform for a black hole of mass $M = 50 \sol{M}$
and dimensionless angular momentum $a = 0.3$.}
\label{fig7}
\end{figure}

\begin{figure}
\begin{center}
\begin{picture}(0,0)
\put(-10,30){\rotatebox{90}{\(2 \Delta \phi( t ) / ( h \omegar )\ (\milli
\second)\)}}
\put(100,-5){time (\(\milli\second\))} 
\end{picture}%
\includegraphics{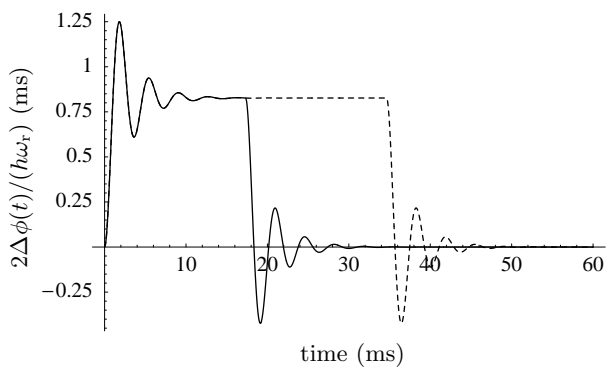}
\end{center}
\caption{The interferometer's response to the ring-down waveform for two
choices of baseline:  $\unit{6000}{\kilo \metre}$ (solid curve) and
$\unit{12000}{\kilo \metre}$ (dashes).  These baselines have light-crossing
times of $\unit{20}{\milli \second}$ and $\unit{40}{\milli \second}$
respectively.}
\label{fig8}
\end{figure}

A detailed exploration of the response of the radio interferometer to the
full parameter space of these ring-down gravitational waveforms is outside
the scope of this document.  To illustrate the nature of the
interferometer's response, we consider the single choice of parameters
consisting of \(M = 50 \sol{M}\), and \(a = 0.3\).  Such a black hole's
fundamental quadrupolar mode has a centre frequency of \(\omega_{\text{s}}
= \unit{1736}{\radian / \second}\), and the ring-down waveform has a
quality factor of \(Q = 2.348\).  The ring-down waveform is shown in Figure
\ref{fig7}.
The interferometer's response to this type of gravitational wave is
depicted in Figure \ref{fig8} which is a plot of \(2 \Delta \phi( t ) / ( h
\omegar)\) as a function of time when the gravitational wave is approaching
from the optimum direction with the optimum polarization.
For these ring-down waveforms, the output of the interferometer is not
proportional to the gravitational wave signal, as it is with monochromatic
gravitational waves, so there is no clear way to define a quantity similar
to the ``effective phase length.''  Comparing to the analysis of
monochromatic gravitational waves in the previous section, the RMS time
average of the function plotted in Figure \ref{fig8} is algebraically equal
to the efficiency times the baseline, \(\efficiency 2 l\) (hence the choice
of milliseconds for the units of the vertical axis).  The RMS average over
infinite time, however, would give a result of 0, so for gravitational
waves localized in time this procedure is not a good way to assess the
strength of the interferometer's output.

The interferometer's response is plotted for two choices of the
interferometer's baseline: \(\unit{6000}{\kilo \metre}\) and
\(\unit{12000}{\kilo \metre}\) (light crossing times of \(\unit{20}{\milli
\second}\) and \(\unit{40}{\milli \second}\) respectively).  The shape of
the interferometer's response as a function of time can be understood as
follows.  Prior to \(t = \unit{0}{\second}\), the gravitational wave has
not yet encountered the interferometer and there is no differential phase
seen.  At \(t = \unit{0}{\second}\), the ring-down wave train wave crosses
the line of sight of the first receiving element, resulting in that station
accumulating a phase change in its received radio signal.  There is then a
constant differential phase between the stations until the wave train
crosses the line of sight of the second receiving element (a little less
than either \(\unit{20}{\milli \second}\) or \(\unit{40}{\milli \second}\)
later).  At that time the second station accumulates the same change in the
phase of its received radio signal and the differential phase between the
stations returns to 0.

\section{Noise in the Interferometer}

To assess the viability of the type of gravitational wave detector
described in this paper, it is necessary to investigate the noise that must
be anticipated in the detector's output.  A full analysis of the noise
level requires a complete specification not only of the detection apparatus
but also of the observation strategies and data analysis techniques to be
used in the extraction of a gravitational wave signal from the detector's
raw output.

The technique of gravitational wave detection described in this paper can,
in principle, be implemented with both optical and radio technologies.
Although it has been implied throughout the document that it is radio
systems that are being considered, the analysis up to this point is
applicable to both.  Optical systems benefit from operation at higher
angular frequencies but radio systems, at least terrestrially, benefit from
much longer baselines and the ability to record the electromagnetic signals
for detailed off-line analysis.  Another advantage of the implementation of
the detector using radio technology is the possibility of simultaneous
multi-use of the technology for other scientific research.  We now
specifically assume a radio-based system.  There will be little discussion
of observation strategies or data analysis techniques as these are beyond
the scope of this paper.

Gravitational wave detection proves to be one of the most challenging
measurement problems in modern physics.  All detection techniques currently
being used or considered require extraordinary technological achievements.
The technique described here does not provide a ``magic bullet'' for
circumventing any of the difficulties in gravitational wave detection so
one must expect that the practical implementation of this technique will
also require an apparatus with extraordinary operating parameters.

\subsection{Thermal Noise in the Electronics}

\begin{figure}
\begin{center}
\begin{picture}(0,0)%
\includegraphics{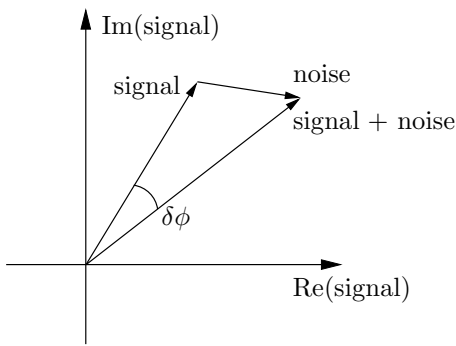}%
\end{picture}%
\setlength{\unitlength}{3947sp}%
\begingroup\makeatletter\ifx\SetFigFont\undefined%
\gdef\SetFigFont#1#2#3#4#5{%
  \reset@font\fontsize{#1}{#2pt}%
  \fontfamily{#3}\fontseries{#4}\fontshape{#5}%
  \selectfont}%
\fi\endgroup%
\begin{picture}(2124,2124)(-11,-1273)
\put(1801,-961){\makebox(0,0)[lb]{\smash{\SetFigFont{10}{12.0}{\familydefault}{\mddefault}{\updefault}Re(signal)}}}
\put(601,689){\makebox(0,0)[lb]{\smash{\SetFigFont{10}{12.0}{\familydefault}{\mddefault}{\updefault}Im(signal)}}}
\put(1802, 88){\makebox(0,0)[lb]{\smash{\SetFigFont{10}{12.0}{\familydefault}{\mddefault}{\updefault}signal + noise}}}
\put(977,-512){\makebox(0,0)[lb]{\smash{\SetFigFont{10}{12.0}{\familydefault}{\mddefault}{\updefault}\(\delta\phi\)}}}
\put(1802,388){\makebox(0,0)[lb]{\smash{\SetFigFont{10}{12.0}{\familydefault}{\mddefault}{\updefault}noise}}}
\put(677,313){\makebox(0,0)[lb]{\smash{\SetFigFont{10}{12.0}{\familydefault}{\mddefault}{\updefault}signal}}}
\end{picture}
\end{center}
\caption{Complex signal and noise phasors for the radio signal being
received by the interferometer.}
\label{fig2}
\end{figure}

The antenna structures, radio receivers and data acquisition systems used
in astronomical radio interferometry applications are not noise-free, and
contribute to the uncertainty in the measurement of the interferometer
phase.  Figure \ref{fig2} shows an example complex phasor diagram for the
radio signal being received and the thermal noise added to it by the
interferometer.
If the signal-to-noise ratio is greater than 1, then it can be seen that
the phase error introduced by the noise is
\begin{equation*}
\delta \phi
	\sim \frac{\magnitude{\text{noise}}}{\magnitude{\text{signal}}}
	= \radio{\snr}^{-1}
\end{equation*}
where the subscript r on the right-hand side identifies this as the radio
interferometer's signal-to-noise ratio.  If we take our gravitational wave
signal to be the time-averaged RMS differential phase in \eqref{eqn7} and
compare this to the phase noise from above, we get
\begin{equation*}
\gw{\snr}
   = \frac{\Delta \rms{\phi}}{\delta \phi}
   = \Delta \rms{\phi} \, \radio{\snr}
\end{equation*}
as the signal-to-noise ratio on a single determination of the differential
phase due to the presence of a gravitational wave.  Since \(\Delta
\rms{\phi} \propto h \frac{\omegar}{\omegagw}\) is typically much smaller
than 1, we see immediately that we are going to need a very high
signal-to-noise ratio in the interferometer in order to see gravitational
waves above the thermal noise.

The signal-to-noise ratio of a two-element radio interferometer is given
by \cite[page 158]{thompson1991}
\begin{equation*}
\radio{\snr}
   = \frac{1}{\sqrt{2 \pi}} \sqrt{\Delta \omegar \Delta t}
   \sqrt{\frac{{\Tant}_{1} {\Tant}_{2}}{{\Tsys}_{1} {\Tsys}_{2}}}
\end{equation*}
where \(\Delta \omegar / (2 \pi)\) is the bandwidth of the receiving system
in Hertz, \(\Delta t\) is the coherent integration time in seconds,
\({\Tant}_{i}\) is the antenna (signal) temperature in Kelvin at the
\(i^{\text{th}}\) antenna, and \({\Tsys}_{i}\) is the system (noise)
temperature in Kelvin at the \(i^{\text{th}}\) antenna.  The antenna
temperature is given by \cite{kraus1966}
\begin{equation*}
\Tant
	= \frac{10^{-26} \Snu \Aeff}{2 \boltzman}
\end{equation*}
where \(\Snu\) is the spectral flux density of the observed radio source in
Janskys (\(\unit{1}{\jansky} = \unit{10^{-26}}{\watt \metre^{-2}
\hertz^{-1}}\)), \(\Aeff\) is the antenna's effective aperture area in
square metres, and \(\boltzman = \unit{1.4 \times 10^{-23}}{\joule
\kelvin^{-1}}\) is the Boltzmann constant.  The factor of 2 is due to the
radio receiver being able to observe only one electromagnetic polarization.
For a ``blind search'' observation of gravitational waves with an angular
frequency of \(\omegagw\), our interferometer's coherent integration time
cannot exceed the inverse Nyquist frequency of \(\Delta t = \pi /
\omegagw\).  If we, further, assume the two receiving stations are
identical (same effective apertures, same system temperatures), then
altogether the phase uncertainty due to thermal noise in the radio
interferometer's electronics is
\begin{equation}
\label{eqn6}
\delta \phi_{\text{r}}
   \approx 4 \times 10^{3} \sqrt{\frac{\omegagw}{\Delta \omegar}}
   \frac{\Tsys}{\Snu \Aeff}
\end{equation}
where \(\Snu\) is in \(\jansky\), \(\Tsys\) is in \(\kelvin\), and
\(\Aeff\) is in \(\metre^{2}\).

\subsection{Other Noise Sources}

The thermal noise in the interferometer's electronics limits one's ability
to measure with precision the difference in phase between the signals being
received at each station.  Even given an exact measurement of the
interferometer's differential phase, it is still difficult to attribute it
exclusively to a passing gravitational wave as there are other processes
that contribute to phase differences between the receiving stations.
Examples of such systematic noise sources are:
\begin{itemize}
\item Time-dependant deformations of the interferometer due to, for
example, seismic noise and tidal distortion of the Earth.

\item Variations in the refractive index of the medium through which the
radio waves travel on their way to the receiving stations.  These include
variations in the refractive index of the Earth's atmosphere as well as
possibly interplanetary or interstellar gases (depending on the distance to
the radio source).

\item Thermal distortions of the antennas and the consequent displacement
of the antenna phase centres.

\item Variations in the instrumental signal delays through cables and
electronics at each antenna site.
\end{itemize}

These noise sources, as well as others not mentioned here, affect present
day radio astronomical and geodetic radio interferometric observations.  In
general it will be necessary to either remove the effects of systematic
noise sources by clever observing strategies and/or sophisticated signal
processing techniques, or to provide calibration systems that are capable
of accounting for their presence.  In some cases, present-day radio
interferometric techniques may be capable of providing the necessary tools.
For example, dispersive effects of the ionosphere and the
interplanetary/interstellar medium can be removed by multi-frequency
observations and radio interferometry phase calibration (pcal) systems
\cite[page 290]{thompson1991} can remove the effects of variations in
instrumental phase delay between stations at the picosecond level
\cite{cannon2003}.

Differential observing strategies may possibly be developed that are
effective in removing systematic effects in the output data.  Figure
\ref{fig4} shows that, as a gravitational wave detector, a radio
interferometer is reasonably directional, and the use of antenna arrays and
beam-forming techniques might be employed to implement a gravitational wave
detector with high directivity.  Such a system combined with multi-beam
radio frequency antennas could, in principle, carry out a differential
search for gravitational waves incident from several directions on the sky
simultaneously.  Differential observing strategies such as this are capable
of cancelling systematic noise sources to a high precision.

The sources of noise mentioned above produce differential phase in the
interferometer through means other than the curvature of space.  Even when
all such noise can be removed from the interferometer's output, leaving
only curvature-induced differential phase, there are sources of curvature
other than gravitational waves.  Phenomena such as seismic and atmospheric
activity can produce changes in local gravity comparable to the expected
amplitude of astrophysical gravitational waves \cite{schutz1999}.  Unlike
ground-based laser interferometers, the radio signals used by the type of
detector investigated here do not travel along geodesics that are entirely
local to the Earth.  For this reason, it is not yet clear the extent to
which such sources of noise contribute to the detector's output but it is
possible this type of detector is less sensitive to them than are entirely
ground-based systems.

A full analysis of the noise from these and other sources and the extent to
which these can be subtracted from the detector's output will be the
subject of future work.

\section{Practical Gravitational Wave Detection}

If we assume for the moment that all systematic sources of differential
phase can be exactly accounted for and subtracted from the detector's
output, then that leaves the interferometer's thermal noise as the only
impediment to the operation of this type of detector.  While this is an
unrealistic situation, it does represent a sensitivity limit which cannot
be improved upon through any amount of creative observation strategies or
signal extraction techniques (with the exception of extending the
interferometer's integration time, and this will be discussed below).

For good gravitational wave signal to noise, it has been established that
long baselines, high frequency radio observations, low receiver noise
temperatures and bright radio sources are required.  In addition, one can
consider performing multiple simultaneous observations of a gravitational
wave:  by constructing an interferometer with more than two elements, the
differential phase across multiple baselines can be measured.  If these
measurements are time-shifted appropriately and summed together, the
thermal noise in the individual measurements will sum incoherently while
any signal that is present will be summed coherently.  For \(N\) baselines,
the relative phase noise could be reduced by as much as a factor of
\(\sqrt{N}\).

\begin{figure}
\begin{center}
\begin{picture}(0,0)
\put(-10,60){\rotatebox{90}{\(h_{\text{min}}\)}}
\put(55,-5){Gravitational wave frequency (\(\milli \hertz\))} 
\end{picture}%
\includegraphics{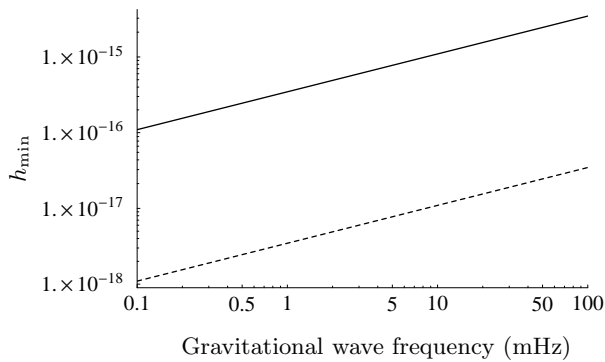}
\end{center}
\caption{The minimum shear required in order to achieve a signal to noise
ratio of 1 in the observation of gravitational waves using a radio
interferometer array with the parameters listed in the text.  Two radio
source spectral flux densities are considered:  $\unit{10^{2}}{\jansky}$
(solid line) and $\unit{10^{4}}{\jansky}$ (dashed line).  Compare to Figure
4 of \cite{cutler2002}.}
\label{fig9}
\end{figure}

Setting \(\Delta \rms{\phi}\) in \eqref{eqn7} equal to \(\delta
\phi_{\text{r}}\) in \eqref{eqn6} and solving for \(h\) gives us the
gravitational wave shear needed to achieve a signal-to-noise ratio of at
least 1.  This is
\begin{multline}
h_{\text{min}}
   = 6 \times 10^{3} \sqrt{\frac{\omegagw^{3}}{\omegar^{2} \Delta \omegar}}
   \frac{\Tsys}{\Snu \Aeff} \,\cdot \\ \magnitude{ ( 1 + \cos \theta ) \cos
   2 \alpha \sin ( \omegagw l \sin \theta \sin \phi ) }^{-1}.
\end{multline}
Optimistic yet achievable numbers for some of the parameters in this
expression are:  a radio observation centre frequency of \(\omegar =
\unit{10^{11}}{\radian / \second}\), a bandwidth of \(\Delta \omegar =
\unit{10^{9}}{\radian / \second}\), system temperatures of \(\Tsys =
\unit{20}{\kelvin}\), effective apertures of \(\Aeff =
\unit{10^{3}}{\metre^{2}}\), and an interferometer consisting of 32
stations for a total of 496 baselines of about \(2 l =
\unit{2000}{\kilo\metre}\) (\(l = \unit{3}{\milli \second}\)) each.  Given
these parameters, Figure \ref{fig9} shows plots of \(h_{\text{min}}\) for
radio source spectral flux densities of \(\unit{10^{2}}{\jansky}\) and
\(\unit{10^{4}}{\jansky}\) (at \(\unit{16}{\giga\hertz}\)).
Flux densities of this order are available naturally from galactic water
vapour maser sources at \(\unit{22}{\giga \hertz}\) or could be provided
artificially, for example by means of a spacecraft-born radio transmitter.
The Canadian Large Adaptive Reflector (CLAR), embodies a design for a
low-cost, large aperture (\(\unit{300}{\metre}\) diameter or larger),
fully-steerable, multi-beam radio astronomy antenna \cite{cote2002,
dewdney2000, cannon1998}.  One technical implementation for the Square
Kilometre Array (SKA) envisages the construction of an array of CLAR
antennas.  Such an interferometer array might, potentially, have use in the
search for gravitational radiation although its anticipated operating
parameters differ somewhat from the numbers quoted above.

To compare the sensitivity of this type of detector to those of other
types, as well as to the strengths of anticipated sources of gravitational
waves, see the recent reviews found in \cite{cutler2002} and
\cite{schutz1999}.  The noise curves in Figure \ref{fig9} are well above
any anticipated gravitational wave signals.  It must be recalled, however,
that these noise curves are for the raw detection of a gravitational wave
signal at or above the noise level: a pen on a chart recorder marking out
the gravitational wave shear as a function of time.  This picture is not
what is meant when it is said that the current generation of gravitational
wave observatories will ``observe'' gravitational waves.  Rather than
obtaining a direct read-out of the gravitational wave shear as a function
of time, one convolves the output of the detector against a template
waveform and compares the correlation coefficient to a baseline coefficient
obtained by correlating the same template against a model of the detector's
noise.  By this method, one extends the coherent integration time of the
interferometer and reduces the output to the probability that the given
waveform is present in the data.  In the case of many astrophysical
sources, such as binary neutron star systems, integration times of months
or even years are anticipated \cite{finn2000}.  This is to be contrasted to
the integration times of \(\leq \unit{10^{3}}{\second}\) that were assumed
for the computation of Figure \ref{fig9}.  Since the noise curves in Figure
\ref{fig9} are inversely proportional to the square root of the integration
time, then if such long integration times could be used in the radio
interferometer one could potentially decrease \(h_{\text{min}}\) by two or
three orders of magnitude.  In the frequency band shown in Figure
\ref{fig9}, such a reduction in the noise levels would make this type of
detector competitive with the LISA detector's noise level as shown in
Figure 4 of \cite{cutler2002}.

It must be stressed, however, that systematic noise has been ignored.  An
analysis of the contribution from systematic noise is complicated by the
availability of sophisticated calibration techniques that can subtract a
great deal of it.  A complete analysis of the influence of systematic noise
will involve not only an estimate of the raw contribution from these
sources but also the development and subsequent analysis of the
effectiveness of observation strategies and data processing techniques.
This will be the subject of future work.

\section{Conclusions}

The technique, introduced here, of using the time-dependent phase
difference between two radio receiving stations to watch for the passage of
a gravitational wave would seem to be a promising method of gravitational
wave observation worthy of further investigation.  In the future, a more
detailed analysis of realistic interferometer configurations and data
analysis techniques will shed more light on true sensitivity of this type
of detector.

\bibliography{references}

\end{document}